\documentclass[aps,floatfix,prd,nofootinbib,superscriptaddress,reprint,showpacs
  ,10pt,preprintnumbers,longbibliography]{revtex4-1}
\usepackage[utf8]{inputenc}
\usepackage[pdftex]{graphicx}
\usepackage{float}
\usepackage{amsmath}
\usepackage{amssymb}
\usepackage{mathtools}
\usepackage{braket}
\usepackage{siunitx}
\usepackage{amsfonts}
\usepackage{dsfont}
\usepackage{array}
\usepackage{bm}
\usepackage{mathrsfs}
\usepackage{pifont}
\usepackage{multirow}
\usepackage{upgreek}
\usepackage[dvipsnames]{xcolor}
\usepackage[pdftex,
  pdftitle={},
  pdfauthor={},
  bookmarks,
  colorlinks,
  linkcolor=myblue,
  citecolor=mymagenta,
  menucolor=black,
  urlcolor=myblue,
  plainpages=false,
  pdfpagelabels,
  hypertexnames=false]{hyperref}
\usepackage{verbatim}
\usepackage{slashed}
\usepackage{cleveref}
\usepackage{ucs}
\usepackage{subcaption}
\usepackage{csquotes}
\usepackage{bbold}

\makeatletter
\let\newfloat\newfloat@ltx
\makeatother
\usepackage{algorithm}
\makeatletter
\renewcommand{\ALG@name}{Algorithm }
\makeatother
\usepackage[noend]{algpseudocode}

\DeclareMathOperator{\im}{i}

\newcommand{\md}{\ensuremath{\mathrm{d}}}

\definecolor{mymagenta}{RGB}{200, 0, 100}
\definecolor{myblue}{RGB}{45, 48, 146}

\graphicspath{{plots/}}

\begin{document}
\title{A Comprehensive Stress Test of Truncated Hilbert Space Bases\\
	against Green's function Monte Carlo in U$(1)$ Lattice Gauge Theory}

\author{Timo Jakobs}
\affiliation{Helmholtz-Institut f\"ur Strahlen- und Kernphysik, University of
  Bonn, Nussallee 14-16, 53115 Bonn, Germany}
\affiliation{Bethe Center for Theoretical Physics, University of Bonn,
  Nussallee 12, 53115 Bonn, Germany}

\author{Marco Garofalo}
\affiliation{Helmholtz-Institut f\"ur Strahlen- und Kernphysik, University of
  Bonn, Nussallee 14-16, 53115 Bonn, Germany}
\affiliation{Bethe Center for Theoretical Physics, University of Bonn,
  Nussallee 12, 53115 Bonn, Germany}

\author{Tobias Hartung}
\affiliation{Northeastern University - London, Devon House, St Katharine Docks,
  London, E1W 1LP, United Kingdom}
\affiliation{Khoury College of Computer Sciences, Northeastern University,
  \#202, West Village Residence Complex H, 440 Huntington Ave, Boston, MA
  02115,
  USA}

\author{Karl Jansen}
\affiliation{Computation-Based Science and Technology Research Center,
  The Cyprus Institute, 20 Kavafi Street, 2121 Nicosia, Cyprus}
\affiliation{Deutsches Elektronen-Synchrotron DESY, Platanenallee 6, 15738
  Zeuthen, Germany}

\author{Paul Ludwig}
\affiliation{Helmholtz-Institut f\"ur Strahlen- und Kernphysik, University of
  Bonn, Nussallee 14-16, 53115 Bonn, Germany}
\affiliation{Bethe Center for Theoretical Physics, University of Bonn,
  Nussallee 12, 53115 Bonn, Germany}

\author{Johann Ostmeyer}
\affiliation{Helmholtz-Institut f\"ur Strahlen- und Kernphysik, University of
  Bonn, Nussallee 14-16, 53115 Bonn, Germany}
\affiliation{Bethe Center for Theoretical Physics, University of Bonn,
  Nussallee 12, 53115 Bonn, Germany}

\author{Simone Romiti}
\affiliation{Institute for Theoretical Physics, Albert Einstein Center for
  Fundamental Physics, University of Bern, CH-3012 Bern, Switzerland}

\author{Carsten Urbach}
\affiliation{Helmholtz-Institut f\"ur Strahlen- und Kernphysik, University of
  Bonn, Nussallee 14-16, 53115 Bonn, Germany}
\affiliation{Bethe Center for Theoretical Physics, University of Bonn,
  Nussallee 12, 53115 Bonn, Germany}
\date{\today}

\begin{abstract}
  A representation of Lattice Gauge Theories (LGT) suitable for simulations
  with tensor network state methods or with quantum computers requires
  a truncation of the Hilbert space to a finite dimensional
  approximation. In particular for U$(1)$ LGTs, several such
  truncation schemes are known, which we compare with each other using
  tensor network states. We show that a functional
  basis obtained from single plaquette Hamiltonians -- which we call
  plaquette state basis -- outperforms the other schemes in two
  spatial dimensions for plaquette,
  ground state energy and mass gap, as it is delivering accurate results
  for a wide range of coupling strengths with a
  minimal number of basis states. We also show that this functional basis can
  be efficiently used in three spatial dimensions.
  Green's function Monte Carlo appears
  to be a highly useful tool to verify tensor network states
  results, which deserves further investigation in the future.
\end{abstract}

\maketitle

\section{Introduction}

Hamiltonian formulations of lattice gauge theories (LGTs) offer promising avenues to address previously intractable challenges, such as real-time evolution and finite density simulations in strongly coupled quantum chromodynamics. 
The rapid advancements in quantum computing (QC) hardware~\cite{DiMeglio:2023nsa} and tensor network (TN) algorithms~\cite{Magnifico:2024eiy} indicate that these goals may soon be achievable.

A priori, the standard Kogut-Susskind Hamiltonian~\cite{Kogut:1974ag} acts on an infinite dimensional Hilbert space.
More specifically, each individual gauge link is already infinite dimensional.
Physical quantum hardware as well as tensor networks require a projection of this space onto finite available resources.
Thus, in order to investigate lattice gauge theories in the Hamiltonian
formalism, the Hilbert space of a gauge link needs to be
digitised. In
other words, a discrete basis needs to be devised such that its truncation efficiently
approximates the eigenstates of the Hamiltonian in the target,
infinite dimensional Hilbert space.

Finding such a basis is complicated by Gauss law requirements and the
involved interplay between the kinetic (or electric) and the potential
(or magnetic) part of the Hamiltonian: the former
dominates for large coupling, the latter for small coupling.
In consequence, for large coupling it
is efficient to work in an eigenbasis of the momentum operators and
apply a cut-off at a certain momentum value.
This approach yields the electric basis or, more generally, the
Clebsch-Gordan expansion~\cite{Zohar:2014qma}. 
For small coupling, on the other hand, there exists an efficient magnetic basis in Abelian
U$(1)$ gauge theories~\cite{PhysRevD.99.074502,Haase2021}, but there
is nothing equivalent know for non-Abelian SU$(N)$ gauge theories.

Rather than considering the extreme regimes of the Hamiltonian, one
can also choose to discretise the gauge group manifold
itself~\cite{Hartung:2022hoz,Jakobs:2023lpp,Jakobs:2025rvz}. 
This ansatz is equivalent to choosing a maximally localised functional
basis, that is a finite number of Dirac $\delta$ functions. Of course,
other functional bases can, in principle, be valid as well.

Using ideas from Ref.~\cite{Bauer:2023jvw} and
Ref.~\cite{PhysRevD.107.074504}, we have developed\footnote{The
fundamental idea was first publicly presented in August 2024 at
Lattice  2024 at The University of Liverpool~\cite{timo:lattice24},
Liverpool, UK.} 
a functional basis that interpolates smoothly between both regimes.
The form of the functional basis used in this paper was first
published as a preprint in September 2024 in Ref.~\cite{Fontana:2024rux}. 
It is obtained using the eigenstates of the single plaquette
Hamiltonian as a basis. Here, we present the explicit results for
U$(1)$ and SU$(2)$, but the concept generalises easily to other gauge
groups. While completing this work, we learned at a workshop at
ECT*~\cite{ect*2025} of the work published now as a preprint in 
Ref.~\cite{Miranda-Riaza:2025fus}, with similar developments to ours
in U$(1)$.
Consistently with the observations in Ref.~\cite{Miranda-Riaza:2025fus}, we find
the single plaquette basis to be very efficient not only in the small coupling limit.
In our U$(1)$ and SU$(2)$ simulations, the proposed basis allowed us to work
with system sizes significantly larger than previously feasible in
tensor network simulations. Even three spatial dimensions can be
approached.

In this work, we present an overview over the different bases.
We benchmark them for multiple system sizes and a vast range of
coupling strengths using tensor network states. 
The results are verified using exact diagonalisation and Green's
function Monte Carlo whenever possible.
In addition to the ground state energy and plaquette expectation
value, we in particular compute the mass gap for a wide range of
coupling-values, providing a considerably less forgiving stress test. 

These developments pave the way towards a much more faithful treatment
of gauge fields in tensor network or quantum computing simulations,
and eventually in first applications like e.g.\ Refs.~\cite{Gross:2025qae,Crippa:2024cqr,Kane:2025ybw,Rosanowski:2025nck,Crippa:2024hso,Meth:2023wzd}. 

The rest of this paper is structured as follows: In
Section~\ref{sec:theory}, we introduce the Hamiltonian formulation of lattice gauge
theories. Then the different digitisation schemes are introduced first
for U(1) in Section~\ref{sec:U(1)} and then for SU(2) in
Section~\ref{sec:SU(2)}. Our results are presented in
Section~\ref{sec:results} and we conclude in
Section~\ref{sec:conclusion}. 
 
\section{Theory}
\label{sec:theory}

The Hamiltonian formulation of lattice gauge theories was first proposed by
Kogut and Susskind in 1974~\cite{Kogut:1974ag}. Similar to Wilson's famous
Lagrangian formulation~\cite{Wilson:1974sk} it is also defined on a cubical
lattice. Contrary its Lagrangian counterpart, only the spatial dimensions are
discretised, while time is kept continuous. The gauge degrees of freedom are
implemented as links connecting the spatial lattice sites. Each link is
classically described by a colour matrix $U$, typically in the fundamental
representation of the gauge group $G$.

Quantum states of the system are described by a many body wave function
\begin{equation}
    \psi\left(\dots, {U_{\mathbf{x}, k}}, \dots \right): G^{N_{\textrm{links}}}
    \, \,
    \rightarrow
    \, \, \mathbb{C} \, ,
\end{equation}
assigning a complex probability amplitude to every classical configuration
$\left\{U_{\mathbf{x}, k}\right\}$ of the gauge links.
Directions and positions of each link are labelled by coordinate
$\mathbf{x}$ and direction $k$.

\subsection{Operators}

We begin by defining the position and momentum operators of the theory.
Similar to regular quantum mechanics the position or link operators modify
the wave function, by multiplication with the respective gauge degree of
freedom.
More precisely

\begin{equation}
    \hat{U}_{\mathbf{x}, k} \, \psi = U_{\mathbf{x}, k}\
    \psi\left(\dots, {U_{\mathbf{x}, k}}, \dots\right) \, .
\end{equation}
From this the plaquette operator can be constructed. It is defined as
\begin{equation}
    \label{eq:plaquette}
    \hat{P}_{\mathbf{x}, i j} =\ \hat{U}_{\mathbf{x},i}\,
    \hat{U}_{\mathbf{x}+a\hat{\mathbf{i}}, j}\,
    \hat{U}^\dagger_{\mathbf{x}+a\hat{\mathbf{j}}, i}\,
    \hat{U}^\dagger_{\mathbf{x},j} \,
\end{equation}
and can be understood as the oriented loop around a fundamental square of the
lattice in the $i$ $j$ plane, starting and ending at position $\mathbf{x}$.

The momentum operators take the shape of Lie
derivatives on the gauge group.
In non-Abelian gauge groups there are left and right momentum operators
$\hat{L}^c_{\mathbf{x},k}$ and $\hat{R}^c_{\mathbf{x},k}$ defined
as
\begin{align}
    \hat{L}^c_{\mathbf{x},k} \, \psi \	 & = -\im
    \frac{\mathrm{d}}{\mathrm{d}\beta}\,
    \psi \left(\dots, e^{- \im \beta \tau_c}\, U_{\mathbf{x},k},
    \dots\right)|_{\beta = 0}\,
    \intertext{and}
    \hat{R}^c_{\mathbf{x},k} \, \psi \	 & = -\im
    \frac{\mathrm{d}}{\mathrm{d}\beta}\,
    \psi \left(\dots,U_{\mathbf{x},k} \, e^{\im \beta
        \tau_c},\dots\right)|_{\beta = 0}\, .
\end{align}
Here $\tau_c$ denote the generators of the Lie algebra in colour
space. In a U$(1)$ Abelian gauge group one has $\hat L=\hat R$.

Similar to quantum mechanics, the link and momentum operators are
characterised by their commutation relations. These are given as
\begin{align}
    [\hat{L}^c_{\mathbf{x},i}, \hat{U}_{\mathbf{y},j}]\  & =\
    -\delta_{\mathbf{x}
    \mathbf{y}} \, \delta_{ij} \, t_c\, \hat{U}_{\mathbf{x},i}\,, \\
    [\hat{R}^c_{\mathbf{x},i}, \hat{U}_{\mathbf{y},j}]\  & =\
    \delta_{\mathbf{x}
        \mathbf{y}} \, \delta_{ij} \, \hat{U}_{\mathbf{x},i} \, t_c \, .
\end{align}
For non-Abelian gauge groups the momentum operators also have non-trivial
commutation relations with each other. These read
\begin{align}\label{eq:L}
    [\hat{L}^a_{\mathbf{x},i}, \hat{L}^b_{\mathbf{y},j}] & = \im f_{abc}
    \, \delta_{\mathbf{x}
    \mathbf{y}} \, \delta_{ij} \, \hat{L}^c \,                           \\
    \label{eq:R}
    [\hat{R}^a_{\mathbf{x},i}, \hat{R}^b_{\mathbf{y},j}] & = \im f_{abc}
    \, \delta_{\mathbf{x}
        \mathbf{y}} \, \delta_{ij} \, \hat{R}^c \, .
\end{align}
Here $f_{abc}$ are the structure constants of the gauge group. The commutator
of left and right momentum operators vanishes.

\subsection{The Hamiltonian}

With these operators we can now define the Kogut-Susskind Hamiltonian. It reads
\begin{equation}
    \label{eq:hamiltonian}
    \begin{split}
        \hat H &=\ \frac{g^2}{2}\sum_{\mathbf{x},c,k}
        \left(\hat{L}_{\mathbf{x},k}^c\right)^2 +
        \frac{N}{g^2}\sum_{\mathbf{x},j<i}
        \mathrm{Tr} \left[\,
        \mathbb{1} -
        \mathrm{Re}\,
        \hat{P}_{\mathbf{x}, ij} \right] \, .
    \end{split}
\end{equation}
Here $N$ is the degree or number of colours of the unitary gauge group $G$.
Note that the prefactor in front of the second (magnetic) term is a
convention, with many different choices found across the literature.

The physical Hilbert space of the theory is further restricted by a constraint
referred to as Gauss's law. It states that any physical state $\ket{\psi}$
needs
to satisfy
\begin{equation}
    \hat{G}^c_{\mathbf{x}} \ket{\psi} = \sum_{k} \left(
    \hat{L}^c_{\mathbf{x},k} + \hat{R}^c_{\mathbf{x} -
        a \hat{\mathbf{k}},k} \right) \ket{\psi} = 0 \, .
\end{equation}
It implements local colour charge conservation at each vertex in the lattice.

The first term of the Hamiltonian is commonly referred to as the electric term.
It contains a sum of Laplace-Beltrami
operators $\sum_{c}(\hat{L}^c)^2$ applied
to each gauge degree of freedom of the theory. It is therefore similar
to the kinetic term in a many particle Hamiltonian in quantum mechanics. The
second term is referred to as the magnetic term. It contains the sum over all
plaquettes and thus implements a four particle nearest neighbour interaction.

At small values of the coupling $g^2$ the latter interaction dominates the
dynamics
of the theory and leads to strong entanglement between the four
links of each plaquette. This entanglement makes efficient numerical
simulations in this region extremely difficult.
As the interaction does not
constrain the wave function of any individual gauge link, na\"ive simulations
that
digitise in terms of single links basis, introduce non-localities in
the electric Hamiltonian, and in turn a diverging basis sizes for $g^2
    \rightarrow 0$ to produce accurate results.

\subsection{Dual Formulation}

One strategy to address the non-localities discussed above is to
reparametrise the Hamiltonian, such that
the magnetic term becomes local. For U(1) such formulations exist and
are commonly used \cite{PhysRevD.99.074502,Weber:2013bea,Haase:2020kaj}.
In the following we will go with the rotor formulation presented in
\cite{PhysRevD.99.074502}. Its Hamiltonian is given by
\begin{equation}
    \hat{H}_{\textrm{dual}} = \frac{g^2}{2} \sum_{\mathbf{x},i} \left(
    \hat{L}_{\mathbf{x}} - \hat{L}_{\mathbf{x}+a\hat{\mathbf{i}}} \right)^2 +
    \frac{1}{2 \,g^2} \sum_{\mathbf{x}} \left( 2 -
    \hat{U}_{\mathbf{x}} - \hat{U}_{\mathbf{x}}^{\dagger} \right) \,.
\end{equation}
After this reparametrization no Gauss law constraint is needed.

In the case of non-Abelian gauge groups the situation is less clear. While
candidates for dual Hamiltonians exist
\cite{Bauer:2023jvw,Fontana:2024rux}, they typically lead
to long
range electric interactions. These are problematic for some of the novel
simulation methods proposed for Hamiltonian lattice gauge theories. In quantum
computers such terms would lead to a higher demand in terms of swap gates, and
tensor network methods typically work most efficiently with nearest neighbour Hamiltonians.
At the time of writing it is unclear, whether a dual formulation with only
local electric interaction exists for the non-Abelian case.

\begin{figure}
    \center
    \begin{subfigure}[c]{0.4\columnwidth}
        \includegraphics[width=\textwidth]{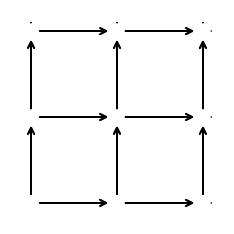}
    \end{subfigure}
    \begin{subfigure}[c]{0.1\columnwidth}
        \scalebox{2}{$\rightarrow$}
    \end{subfigure}
    \begin{subfigure}[c]{0.4\columnwidth}
        \includegraphics[width=\textwidth]{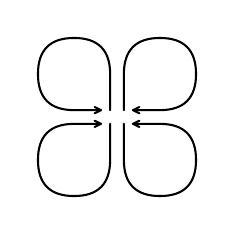}
    \end{subfigure}
    \caption{Sketch of the $3 \times 3$ system in the original formulation on
        the left and the dual formulation on the right.}
    \label{fig:su23x3sketch}
\end{figure}

In the following we will run our non-Abelian tests on a $3\times3$ lattice with
open boundary conditions. This system can be reparametrised in terms of four
plaquette degrees of freedom anchored at the central vertex (see
\cref{fig:su23x3sketch}). Dynamics on the eight remaining string degrees of
freedom are prohibited by Gauss law. The resulting Hamiltonian reads
\begin{equation}
    \begin{split}
        \hat{H}_{3\times3} & = 2 g^2 \sum_{i=1}^{4} \sum_{c=1}^{3}
        \left(
        \hat{L}^c_i \right)^2
        \\
        & \quad + g^2 \sum_{c=1}^{3} \left( \hat{L}^c_1 \,
        \hat{L}^c_2 + \hat{L}^c_3 \, \hat{L}^c_4 +
        \hat{R}^c_1 \, \hat{R}^c_3 + \hat{R}^c_2 \,
        \hat{R}^c_4 \right) \\
        & \qquad \quad \qquad \qquad \qquad \qquad + \frac{2}{g^2} \sum_{i =
            1}^{4} \mathrm{Tr} \left[ \mathbb{1} - \hat{U}_i\right] \, .
    \end{split}
    \label{eq:su23x3Hamiltonian}
\end{equation}
The one remaining Gauss law constraint reads
\begin{equation}
    \hat{G}^c \ket{\psi} = \sum_{i=1}^{4} \left(\hat{L}^c_i +
    \hat{R}^c_i\right) \ket{\psi} = 0 \,.
\end{equation}
In our simulations we will enforce it via the addition of a penalty term
\begin{equation}
    \hat{H}_{\mathrm{penalty}} = \kappa \sum_c \left(\hat{G}^c \right)^2
\end{equation}
where $\kappa$ is a large positive real number.

\section{Digitisation Approaches for U(1)}
\label{sec:U(1)}
To run numerical simulations of the theory we need to first digitise the
Hilbert space of functions
\begin{equation}
    \psi : \, \, G^{N_{\textrm{links}}} \rightarrow \mathbb{C}
\end{equation}
where $G$ is the gauge group of the theory. To do so we begin with a
local basis $\hat{\phi}_n$ for the space of functions
\begin{equation}
    \psi : \, \, G \rightarrow \mathbb{C}
\end{equation}
Then one can take products of these functions to obtain a basis of the full
subspace
\begin{equation}
    \ket{n_1, \dots, n_{N_{\textrm{links}}}} = \prod_{i=1}^{N_{\textrm{links}}}
    \hat{\phi}_{n_i}
\end{equation}
In the following we will present several choices for these basis functions for
U(1) and SU(2). Then we will test them in numerical simulations.

We begin with U(1). To give explicit forms of the single link basis states
first we introduce a parametrization of the group:
\begin{equation}
    U(\varphi) = \mathrm{e}^{\im \varphi} , \quad \varphi \in (-\pi, \pi]\,.
    \label{eq:u1parametrisation}
\end{equation}

\subsection{Electric Basis Operators}

The most common choice of basis functions is given by
\begin{equation}
    \hat{\phi}^m_{\textrm{el.}} (\varphi) = \frac{1}{2 \pi} \mathrm{e}^{\im
        m \varphi}\, , \quad m \in \mathbb{Z} \,.
\end{equation}
In this basis the electric operator is diagonal
\begin{equation}
    \bra{m} L \ket{n} =  m \, \delta_{m,n} ,
\end{equation}
hence it's called the electric basis. The link operator takes the shape of a
ladder operator:
\begin{equation}
    \bra{m} U \ket{n} = \delta_{m,n+1}
\end{equation}
To obtain finite operators one truncates by only considering states for which
$|m| \leq m_{\max}$. This will then give operators of dimension
$d_{\textrm{op}} = 2 m_{\max} + 1$.

As the electric basis functions are the low-lying eigenstates of the electric
Hamiltonian they perform very well at large values of the coupling $g^2$. In
the limit $g^2 \to 0$, i.e. the continuum limit of the theory, an
ever-increasing number of states is needed to achieve accurate results.

\subsection{Magnetic Basis Operators}

One way to address this are the magnetic basis operators proposed in
\cite{Haase2021}. Here first the link operators are modified to be the ones of
a $Z_{2 \ell + 1}$ subgroup of $U(1)$:
\begin{equation}
    U \ket{n} \rightarrow \mathcal{U} \ket{m} = \begin{cases}
        \ket{-m}    & m + 1 = \ell       \\
        \ket{m + 1} & \textrm{otherwise} \\
    \end{cases}
\end{equation}
Then one can define a discrete Fourier-transform in the form of
\begin{equation}
    \mathcal{F} = \frac{1}{\sqrt{2 \ell + 1}} \sum_{\mu,\nu=-\ell}^{\ell} \exp
    \left( \frac{2 \pi \im \mu \, \nu}{2 \ell + 1} \right) \ket{\mu} \bra{\nu}\,.
\end{equation}
Once applying this transformation the link operator becomes diagonal:
\begin{equation}
    \mathcal{F} \mathcal{U} \mathcal{F}^\dagger = \sum_{\mu= -\ell}^{\ell} \exp
    \left( - \frac{2 \pi \im \mu}{2 \ell + 1} \right) \ket{\mu} \bra{\mu}
\end{equation}

To then gain a computational advantage towards weak couplings, one can then
truncate the transformed operator to only contain the states where $\mathcal{F}
    \mathcal{U} \mathcal{F}^\dagger$ is sufficiently close to 1. This is done,
because the wave function is expected to vanish for configurations where
$\mathcal{F} \mathcal{U} \mathcal{F}^\dagger \neq 1$ at weak couplings.

This is equivalent to modifying the Fourier transformation to read
\begin{equation}
    \mathcal{\tilde{F}} = \frac{1}{\sqrt{2 \ell + 1}}
    \sum_{\mu=-n_{\max}}^{n_{\max}} \sum_{\nu=-\ell}^{\ell} \exp
    \left( \frac{2 \pi \im \mu \, \nu}{2 \ell + 1} \right) \ket{\mu} \bra{\nu}\,,
\end{equation}
where $n_{\max} \leq \ell$ and relates to the actual operator dimension again
as $d_{\textrm{op}} = 2 n_{\max}+1$.

To obtain accurate results the integer parameter $\ell$ needs to be tuned for a
given operator dimension and coupling. In the following we pick $\ell$ such
that it minimises the error of the ground state energy of a single plaquette
system.

While more elaborate tuning schemes are certainly conceivable, we will show
that this produced optimal results for the system sizes and observables
considered in this article.

\subsection{Plaquette State Basis}

\begin{figure*}
    \begin{subfigure}[t]{0.48\textwidth}
        \includegraphics[width=0.95\columnwidth]{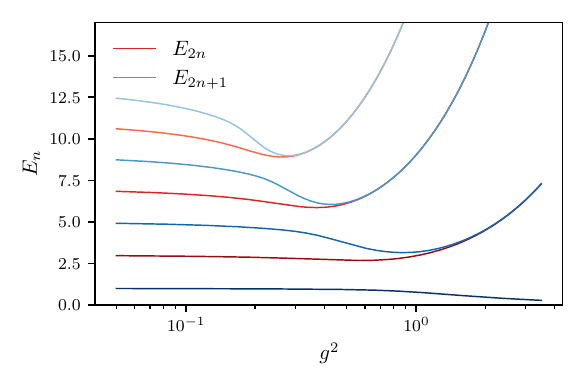}
        \caption{The low energy spectrum of the single plaquette
            system as a function of the coupling $g^2$.}
        \label{fig:u1SpSpectrum}
    \end{subfigure}
    \hfill
    \begin{subfigure}[t]{0.48\textwidth}
        \includegraphics[width=0.95\columnwidth]{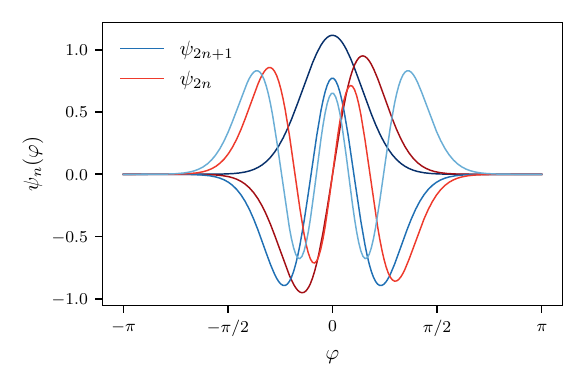}
        \caption{The wave functions corresponding to the single
            plaquette
            eigenstates at a fixed coupling of $g^2=0.1$.}
        \label{fig:u1SpStates}
    \end{subfigure}
    \caption{Spectrum and eigenstates of the single plaquette system.}
    \label{fig:mtOverview}
\end{figure*}

Lastly we will propose a functional basis for U(1) LGTs, which we call
plaquette state basis (see also Ref.~\cite{Miranda-Riaza:2025fus} for
an independent derivation). For this we start
by studying a single plaquette in pure gauge theory. In the dual formulation
the Hamiltonian is given by:
\begin{equation}
    \hat{H}_{\textrm{SP}} = -2 g^2 \frac{\md^2}{\md\varphi_P^2} + \frac{1}{g^2}
    \left( 1 - \cos \varphi_P \right)
\end{equation}
To calculate its eigenstates one has to solve the stationary Schrödinger
equation given by:
\begin{equation}
    \hat{H}_{\textrm{SP}} \ket{\psi_n} = E_n  \ket{\psi_n}
\end{equation}
Plugging in the Hamiltonian we thus end up with a differential equation of the
form
\begin{equation}
    \left( -\frac{\md^2}{\md \varphi^2} + \frac{1}{2 g^4} \left( 1 - \cos
        \varphi
        \right) - \frac{E_n}{2 g^2}
    \right) \psi_n (\varphi) = 0\,.
    \label{eq:u1spdgl}
\end{equation}
This equation is known as Mathieu's differential equation \cite{Mathieu1868}.
In its canonical form it reads
\begin{equation}
    y(x)'' + \left( \lambda - 2 q \cos (2x) \right) y = 0\,.
\end{equation}
To bring \cref{eq:u1spdgl} into this form we first need to reparametrise
$\varphi \rightarrow 2 x $. Then we can read of
\begin{equation}
    q = -\frac{1}{g^4} \quad \textrm{and} \quad \lambda =  \frac{2 E_n}{g^2}-
    \frac{2}{g^4} \,.
\end{equation}
The solutions of this equation are periodic solution, whenever $\lambda$ is
equal to one of the Mathieu characteristic numbers $a_n(q)$ or $b_n(q)$. Due to
the periodicity of the parametrization of the gauge group, only periodic
solutions are of interest.
Thus, the spectrum of the one plaquette system is given by
\begin{equation}
    E_n = \frac{\lambda \, g^2}{2} + \frac{1}{g^2} \, .
\end{equation}
The corresponding eigenfunctions are denoted as $\mathrm{se}_n$
and $\mathrm{ce}_n$ and are called the sine- and cosine-elliptic functions
respectively. They can be classified by periodicity and parity according to
\cref{tab:mathieu-solutions}.
\begin{table}
    \begin{tabular}{l|c|c}
        Solution             & Parity & Period  \\
        \hline
        $\mathrm{ce}_{2n}$   & even   & $\pi$   \\
        $\mathrm{ce}_{2n+1}$ & even   & $2 \pi$ \\
        $\mathrm{se}_{2n}$   & odd    & $\pi$   \\
        $\mathrm{se}_{2n+1}$ & odd    & $2 \pi$ \\
    \end{tabular}
    \caption{Solutions of the Mathieu equation.}
    \label{tab:mathieu-solutions}
\end{table}
As our parametrization defined in \cref{eq:u1parametrisation} has a period of
$\pi$ (after the reparametrization) only the $\pi$ periodic solutions are valid
eigenstates of the system.

We plot the single plaquette spectrum in \cref{fig:u1SpSpectrum} and show the
corresponding eigenstates at a coupling of $g^2 = 0.1$ in
\cref{fig:u1SpStates}. (Describe this a bit more)

A complete set of basis functions is thus given by
\begin{equation}
    \psi_n (\varphi) = \begin{cases}
        \frac{1}{\sqrt{\pi}} \mathrm{ce}_{n+1} (2 \varphi) & n \,\,\,
        \textrm{odd,}                                                  \\
        \frac{1}{\sqrt{\pi}} \mathrm{se}_{n} (2 \varphi)   & n \,\,\,
        \textrm{even.}
    \end{cases}
\end{equation}

To obtain a finite basis needed for simulations we once again truncate by
imposing $n \leq 2 n_{\max} + 1$. The operators are obtained by numerically
solving the integrals
\begin{equation}
    \mathcal{O}_{ij} = \bra{\psi_i} \mathcal{O} \ket{\psi_j} =
    \int_{0}^{2 \pi} \md \varphi \, \overline{\psi_i}(\varphi) \, \mathcal{O}
    \, \psi_j (\varphi) \,.
\end{equation}
It is possible that analytical solutions to these integrals exist. However, for
now we will leave their derivation to future researchers.

Note that that squared operators, such as $\hat{L}^2$ were integrated out
separately and not simply taken as the matrix squared of the $\hat{L}$
operator. This is done, because $\hat{L}$ does not preserve the subspace of
basis functions. Thus, we expect more accurate results, when treating
$\hat{L}^2$ as an independent operator.

\section{Digitisation Approaches for SU(2)}
\label{sec:SU(2)}
Next we will show how some of these approaches translate to the non-Abelian
case of SU(2).

\subsection{Clebsch-Gordan Expansion}

Here the equivalent of the electric basis operators was first formulated in
\cite{PhysRevD.91.054506}. The electric eigenstates are labelled by their
eigenvalues with respect to the Laplace Beltrami operator $\sum_c
    (\hat{L}^c)^2$, as well as one of the left and right momentum operators.
Here typically $\hat{L}^3$ and $\hat{R}^3$ are chosen. This gives a set of
states
\begin{equation}
    \begin{split}
        \ket{j, m_L, m_R},
        &\quad j \in \mathbb{N}/2 ,\\
        & \, m_L,m_R \in \{ -j,-j+1,\ldots,j-1,j \}
        \, ,
    \end{split}
\end{equation}
which obey
\begin{align}
    \sum_c \left( \hat{L}^c \right)^2 \ket{j, m_L, m_R} & = j (j+1) \ket{j,
        m_L,
        m_R}
    \\
    \hat{L}^3 \ket{j, m_L, m_R}                         & = m_L \ket{j, m_L,
        m_R}
    \intertext{and}
    \hat{R}^3 \ket{j, m_L, m_R}                         & = - m_R \ket{j, m_L,
        m_R}
    \,
    .
\end{align}
The action of $\hat{L}^1$ and $\hat{L}^2$ can be defined in terms of raising
and lowering operators
\begin{equation}
    \begin{split}
        & \left( \hat{L}^1 \pm	\im \hat{L}^2 \right) \ket{j, m_L, m_R} \\
        & \qquad\qquad = \sqrt{ j (j+1) - m_L(m_L \pm 1)} \ket{j, m_L \pm 1,
            m_R} \, .
    \end{split}
\end{equation}
Similarly, $\hat{R}^1$ and $\hat{R}^2$ obey
\begin{equation}
    \begin{split}
        & \left( \hat{R}^1 \mp	\im \hat{R}^2 \right) \ket{j, m_L, m_R} \\
        & \qquad\qquad = -\sqrt{ j (j+1) - m_R(m_R \pm 1)} \ket{j, m_L , m_R
            \pm 1} \, .
    \end{split}
\end{equation}
Lastly the link operators need to be obtained. This is done via a
Clebsch-Gordan (CG) expansion~\cite{PhysRevD.91.054506}:
\begin{equation}
    \label{eq:U.CG.expansion}
    \begin{split}
        U_{c_1 c_2}  &= \sum_{j, m_L, m_R} \sum_{j{'}, m_L{'}, m_R{'}}
        \sqrt{\frac{2j +1}{2j{'}+1}} \times
        \\
        &
        \braket{j \, m_L , 1/2 \, m_1 | j' \, m_L{'}}
        \braket{j{'} \, m_R{'} | j \, m_R, 1/2 \, m_2}
        \\
        &
        \, \ket{j{'} \, m_L{'} \, m_R{'}} \bra{j \, m_L \, m_R}\, ,
    \end{split}
\end{equation}
where the colour indices $c_1, c_2$ correspond to $m_1, m_2$ according to the
following convention: $m_i = 1/2 - c_i$.

The resulting operators couple states with the main quantum number $j$ to
states with $j{'}=j \pm 1/2$. Overall these operators fulfil all the
previously stated commutations relations exactly. However, similarly to the
electric operators in U(1) they fail to capture the weak coupling regime of the
theory.

\subsection{Plaquette State Basis}

This can again be solved by considering the states of a single plaquette. Its
Hamiltonian reads
\begin{equation}
    \hat{H}_{\textrm{SP}} = 2g^2 \sum_c \left( \hat{L}^c \right)^2 +
    \frac{4}{g^2} \left(1 - \cos\rho\right)\,.
\end{equation}
The angles $\rho$, $\theta$ and $\varphi$ parametrise the SU(2) gauge group,
where a group element $U$ is given by
\begin{align}
    U (\rho, \theta, \varphi) = \cos \rho \, \mathbb{1} - \im \sin \rho \,
    \vec{n}
    (\theta, \varphi) \cdot \vec{\sigma}
\end{align}
with
\begin{align}
    \vec{n} (\theta, \varphi) =
    \begin{pmatrix}
        \sin \theta \cos \varphi \\
        \sin \theta \sin \varphi \\
        \cos \phi
    \end{pmatrix} \, .
\end{align}
In these coordinates $\sum_c \left( \hat{L}^c\right)^2$ takes the shape of the
Laplace-Beltrami operator on the group manifold:
\begin{equation}
    \sum_c \left( \hat{L}^c \right)^2 = -\frac{1}{4 \sin^2 \rho} \left(
    \frac{\partial}{\partial \rho} \left( \sin^2 \rho \,
        \frac{\partial}{\partial \rho}\right) + \Delta_{S_2}\right)
\end{equation}
Here $\Delta_{S_2}$ is the Laplace Beltrami Operator on the 2-sphere in
terms of the spherical coordinates $\theta$ and $\varphi$. Therefore, we can
make a separation Ansatz:
\begin{equation}
    \phi_n(\rho, \theta, \varphi) = \frac{u(\rho)}{\sin \rho} Y(\theta,
    \varphi)
\end{equation}
The spherical part is solved by the spherical Harmonics $Y_{lm}$ which fulfil
\begin{equation}
    \Delta_{S_2} Y_{lm} = l (l+1) Y_{lm} \, .
\end{equation}
This then leads to the radial equation given by 
\begin{equation}
    u'' + \left(1 - \frac{8}{g^4} \left(1 - \cos \rho \right) +
    \frac{2 E_n}{g^2} + \frac{l (l+1)}{\sin^2 \rho} \right) u = 0 \,.
    \label{eq:su2radial}
\end{equation}
For the gauge invariant case, i.e. $l=0$ this again reduces to the Mathieu
equation. Solutions again need to have a period of $\pi$. Additionally, only
symmetric solutions are differentiable at the poles. As the additional quotient
of $\sin \rho$ itself is anti-symmetric only anti-symmetric solutions of the
Mathieu equations can be considered.

After reparametrization from $\rho \rightarrow 2x$ we can identify
the coefficients of the Mathieu equation to be
\begin{equation}
    q= - \frac{16}{g^4} \qquad \text{and} \qquad \lambda = 4 + \frac{8
        E_n}{g^2} - \frac{32}{g^4} \, .
\end{equation}
With this we can write down a set of orthonormal basis states as
\begin{equation}
    \hat{\phi}_{nlm} (\rho, \theta, \varphi) = \frac{\mathrm{se}_{2n + 2}(q;
        \rho / 2)}{\sin \rho } Y_{lm} (\theta, \varphi) \, .
\end{equation}
Note that while these functions are an orthonormal basis of the Hilbert space,
they are only the eigenfunctions of the single plaquette system for $l=0$.
To fix this one would need to solve \cref{eq:su2radial} separately for $l \neq
    0$. However, to the author's knowledge this can only be done numerically.
The actual set of eigenfunctions is likely to perform much better, as suggested
in \cite{Fontana:2024rux}.

To obtain finite operators we again truncate. In this case this is done by
imposing $l \leq l_{\max}$ and $n \leq n_{\max}$ giving a total operator
dimension of
\begin{equation}
    d_{\textrm{Op}} = (n_{\max} + 1) (l_{\max} + 1)^2 \, .
\end{equation}
The operator matrices are then again obtained via numerical integration.
 
\section{Results}
\label{sec:results}

Next we want to test these operators in numerical simulations. To do
so we use the ITensor DMRG algorithm\cite{itensor}. This is done by mapping
each U(1) rotor in our two-dimensional lattice onto the sites of a one
dimensional MPS. More specifically the rotor found at lattice coordinate $x$
and $y$ is mapped to an MPS site position
\begin{equation}
  \label{eq:indexing}
  i_{\textrm{MPS}} = x (L-1) + y \,.
\end{equation}
This way we will eventually still encounter an exponential slow down, when
increasing the system size in the $y$ direction, but get much better scaling in
the $x$ direction. So while certainly not efficient for larger two-dimensional
systems, it allows us to explore significantly larger system sizes, than with a
regular Eigensolver.

The SVD cutoff parameters and energy convergence criteria in ITensor were tuned
such that we are confident that the finite bond dimension is not contributing a
significant error in the following results. It should thus just be viewed as
more memory efficient Eigensolver for sparse systems with local interactions.
Nevertheless, more appropriate tensor network algorithms will be needed in the
future to explore larger system sizes efficiently.

Furthermore, we will compare the digitised Hamiltonian results to results
obtained from Green's function Monte Carlo (GFMC) Methods. For details on this
algorithm, please refer to \cref{apdx:gfmc}. In short one can obtain good
estimates of the ground state energy and ground state plaquette expectation
value.

\subsection{U(1) Results}

To begin our numerical tests we study a $4 \times 4$ system, with open boundary
conditions in the dual U(1) formulation. This system contains 9 rotor degrees
of freedom. Thus, it should contain enough rotor interactions of the full
theory to give a realistic benchmark, while keeping the overall computational
costs manageable.

\begin{figure}
    \includegraphics[width=0.95\columnwidth]{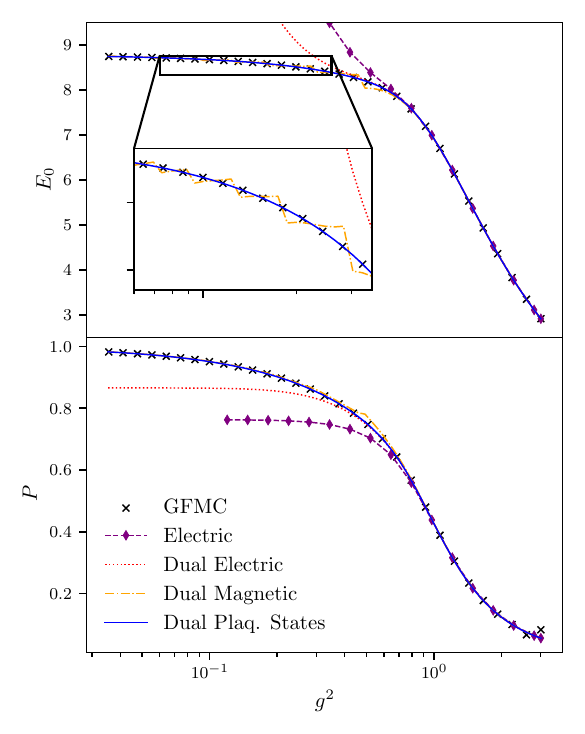}
    \caption{The ground state energy (top) and ground state
        plaquette expectation value (bottom) as a function of the
        coupling $g^2$ for a $4 \times 4$ system with U$(1)$ gauge. This is done
        for the original Kogut-Susskind Hamiltonian using electric basis
        operators (purple), and for the dual formulation using electric (red),
        magnetic (orange) and plaquette state operators (blue). All truncations
        are chosen such that $d_{\mathrm{op}} = 5$. Their results are compared
        to GFMC results shown by the black crosses.}
    \label{fig:u14x4ResultsGnd}
\end{figure}

In \cref{fig:u14x4ResultsGnd} we plot the ground state energy $E_0$ and ground
state plaquette expectation value $P$ as a function of the coupling $g^2$.
Shown are results for the electric basis operators in red, the magnetic basis
operators in orange and the plaquette state operators in blue. For all
operators the truncation was chosen such that $d_{\mathrm{Op}} = 5$.
For comparison, we also include Green's function Monte Carlo (GFMC) results,
plotted as the black crosses, as well as a simulation of the original
Kogut-Susskind Hamiltonian with electric operators also truncated at
$d_{\mathrm{Op}} = 5$.

As expected both simulations using the electric operators quickly diverge from
the GFMC results when decreasing the coupling, while the magnetic and
plaquette state operators stay closely match the GFMC results at all
couplings. As can be seen in the zoomed in panel, the magnetic basis functions
oscillate around the GFMC results. Each jump in this zig-zag pattern
corresponds to a change in the value of $\ell$ when tuning for a given coupling
and operator dimension. The plaquette state operators do not show this
behaviour and thus lead to much smaller deviations from the reference results.
\begin{figure}
    \includegraphics[width=0.95\columnwidth]{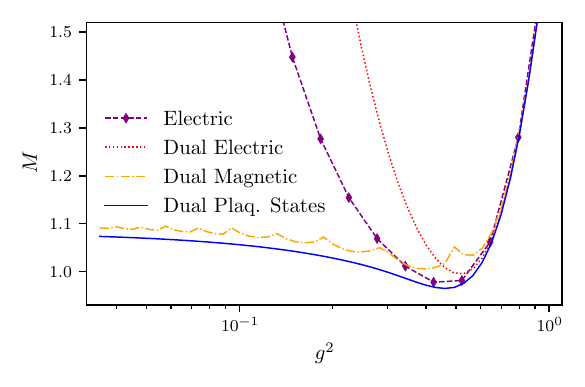}
    \caption{The mass gap $M$ as a function of the coupling $g^2$
        for the same system and operators as \cref{fig:u14x4ResultsGnd}.}
    \label{fig:u14x4ResultsMass}
\end{figure}

Next, we study the mass gap $M = E_1 - E_0$. While it possible to extract
mass gaps from Green's function Monte-Carlo simulations \cite{Hamer2000}, we did
not manage to achieve satisfactory precision in our implementation. Therefore,
we only show the electric simulation of the original Hamiltonian
as well as the three simulations of the dual Hamiltonian in
\cref{fig:u14x4ResultsMass}. The operators are again truncated such that
$d_{\mathrm{Op}} = 5$.

The behaviour of $M$ as a function of the coupling is qualitatively quite
similar to the ground state observables. Both electric formulations again
diverge towards small couplings, while the magnetic basis and plaquette state
basis operators again are stable in the weak coupling limits. The plaquette
states give a marginally smaller mass gap, and the magnetic basis states again
show sharp transitions, when changing $\ell$.

\begin{figure}
    \includegraphics[width=0.95\columnwidth]{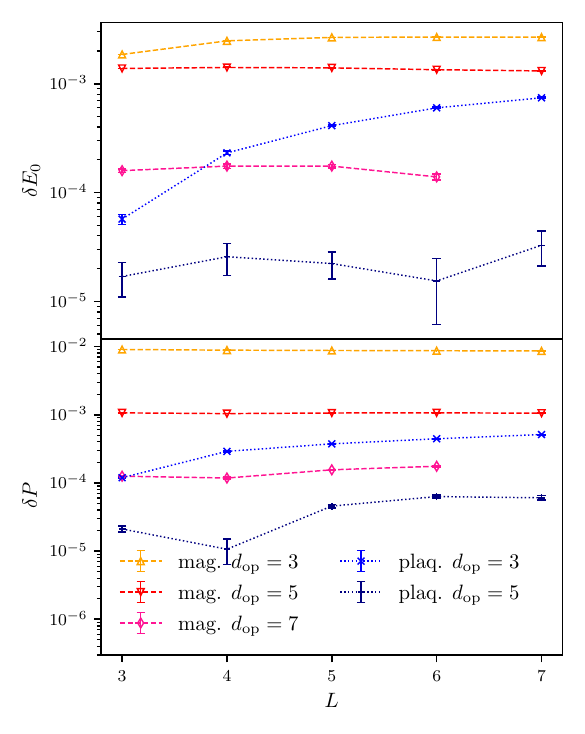}
    \caption{The relative deviations of the ground state energy (top)
    	and the plaquette expectation value (bottom) at a coupling of $g^2=0.1$
        as a function of the system size $L$. This is done for magnetic
        operators at
        three different truncations (orange, red, pink) and for plaquette state
        operators at two different truncations (blue, navy).}
    \label{fig:u14x4Volume}
\end{figure}

To get a more quantitative idea of the accuracy of the different operators we
study the relative deviations from the GFMC results. We define the relative
deviation $\delta \mathcal{O}$ of an observable $\mathcal{O}$ as
\begin{equation}
    \delta \mathcal{O} = \left| \frac{ \mathcal{O}_{\textrm{measured}} -
    \mathcal{O}_{\textrm{ref}}}{\mathcal{O}_{ \textrm{ref}}} \right|\,.
\end{equation}
In \cref{fig:u14x4Volume} we plot the relative deviations with respect to the
GFMC results of $E_0$ and $P$ at a fixed coupling of $g^2=0.1$ as a function
of the lattice size $L$.

This provides insights into how our results for the small system sizes
compare to more substantial volumes. Shown in orange, red and pink are magnetic
operators of with operator dimensions of $3$, $5$, and $7$. The blue and dark
blue points are plaquette state operators of dimension $3$ and $5$.

Similarly to the previous results, the plaquette states show much lower
deviations, than the magnetic operators. For both deviations decrease, when
increasing the operator dimension. The smallest deviation reached by increasing
the operator dimension, seems to be around $10^{-5}$ for both the ground state
energy and the plaquette expectation value. It is likely that the constraining
factor here are unaddressed systematics in the GFMC results. For a more
detailed discussion of these refer to \cref{apdx:gfmc}.

Neither of the operators show a strong volume dependence. The most
pronounced effect is visible for the smallest truncations in both
approaches. Here, the deviations in the energy increase by about one order of
magnitude from the smallest to the largest lattice. For the larger truncations
the deviations are mostly independent of the volume. This leaves us hopeful
that the proposed operators will also perform well in larger simulations.

\begin{figure}
    \includegraphics[width=0.95\columnwidth]{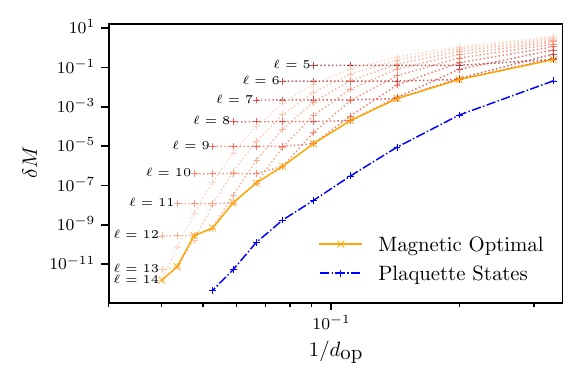}
    \caption{The relative deviation of the mass gap $\delta M$,
        with respect to the result at $d_{\textrm{op}} = 33$, shown as
        a function of the inverse operator dimension $d_{\textrm{op}}^{-1}$ for
        a $4\times4$ system at a coupling of $g^2=0.1$. In blue, we show
        results for the plaquette state operators, while the orange points are
        results obtained from the magnetic operators with optimal $\ell$. The
        cloud of dashed lines above the orange line shows other possible
        choices of the $\ell$ parameter of the magnetic operators.
    }
    \label{fig:u14x4MassConvergence}
\end{figure}

Lastly we want to get an impression of the convergence behaviour of the mass
gap. For this we will compare to a high resolution plaquette state result at
$d_{\textrm{op}} = 33$ to results obtained with smaller truncation at a
coupling of $g^2=0.1$.

The results are depicted in \cref{fig:u14x4MassConvergence}. Here we plot the
relative deviation $\delta M$ of the reference Mass as a function of the
inverse operator dimension for the plaquette state operators in blue.
Additionally, results for the magnetic basis operators are shown. For the
red line $\ell$ was chosen such that the error of the single plaquette ground
state energy is minimised for a given $n_{\max}$. The dotted orange lines show
other possible choices for $\ell$.

Overall the plaquette states show between two and three orders of magnitude
less deviation from the reference result at fixed operator dimension when
compared to the best magnetic basis operators. However, both methods converge
to the reference result, and no significant difference in the convergence rate
can be observed.

Furthermore, the proposed criterion for tuning $\ell$ in the magnetic basis
operators seems to give optimal results. This can be seen, as the orange line
obtained from tuned magnetic basis operators bounds the cloud of all possible
magnetic basis operators from below. It is possible that this only holds at
small system sizes, and different methods are needed for larger systems.

\begin{figure}
    \includegraphics[width=0.95\columnwidth]{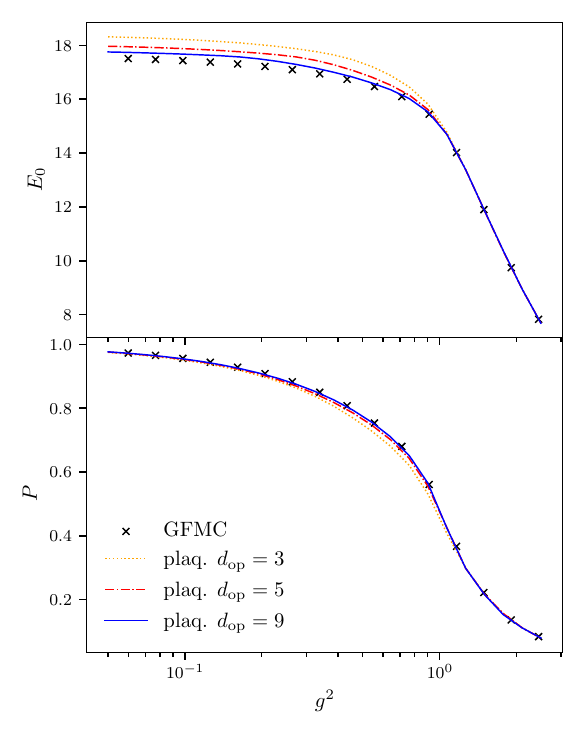}
    \caption{The ground state energy and plaquette expectation
        value (like in fig.~\ref{fig:u14x4ResultsGnd}) for a three-dimensional
        system of size $2 \times 3^2$ as a function of the coupling. Shown are
        results for the plaquette states at three different truncations.}
    \label{fig:u13x3x3GndState}
\end{figure}

Next, we applied the plaquette state basis for simulations of U$(1)$
LGT in three spatial dimensions, using the same snake MPS solver as
before with adapted mapping eq.~\eqref{eq:indexing}. This represents
likely a situation where the limits of snake 
MPS are tested, which is why we could simulate a $2 \times 3^2$ system only.
The corresponding results are shown in \cref{fig:u13x3x3GndState},
where we again compare to GFMC results. Overall deviations seem
significantly larger than in the two-dimensional case for comparable
$d_\mathrm{op}$, which can be understood by the larger number of
interaction terms in the Hamiltonian. Nevertheless, with a small
number of plaquette basis states a range of squared coupling values
$g^2\in[5\cdot10^{-2},2.5]$ can be simulated with small deviations to
the GFMC result. We take this as a strong indication that the
plaquette state basis can enable simulations in three spatial
dimensions without large truncation effects.

\subsection{SU(2) Results}

Finally, we would like to present some results for SU(2). Here we study the
$3\times3$ system described by the Hamiltonian given in
\cref{eq:su23x3Hamiltonian}. Gauss law is enforced via the described penalty
term. $\kappa=10$ proved to be large enough to filter the physical states of
interest. This can be checked, by ensuring that the expectation value of the
penalty Hamiltonian vanishes.

\begin{figure}
    \includegraphics[width=0.95\columnwidth]{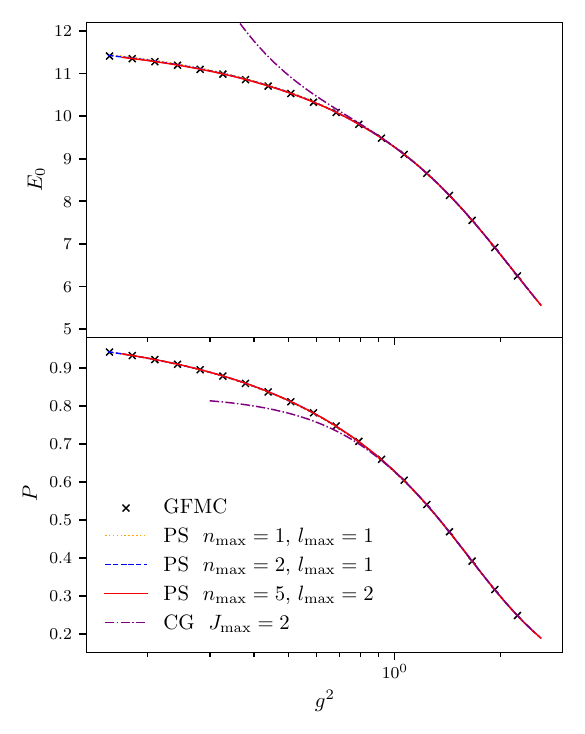}
    \caption{The ground state energy and ground state plaquette
        expectation value as a function of the coupling for a $3\times3$ system
        of SU(2) gauge theory. This is done for the Clebsch-Gordan Operators
        with a truncation of $J_{\max} = 2$ shown in purple and for three sets
        of plaquette state operators at different truncations in orange, blue
        and red respectively.}
    \label{fig:su2-gndState}
\end{figure}

In \cref{fig:su2-gndState} we again plot the ground state energy and plaquette
expectation value as a function of the coupling $g^2$. Shown are results for
Clebsch-Gordan operators truncated at $J_{\max} \leq 2$ ($d_{\textrm{op}} =
    55$)
in purple, and for the plaquette state operators at three different
truncations. They have operator dimensions of $8$, $12$ and $45$ respectively.
Once again we compare them to GFMC results. Similar to the results for U(1),
the Clebsch-Gordan operators eventually diverge towards small couplings, while
the plaquette state operators again match the GFMC results well at all
couplings.

\begin{figure}
    \includegraphics[width=0.95\columnwidth]{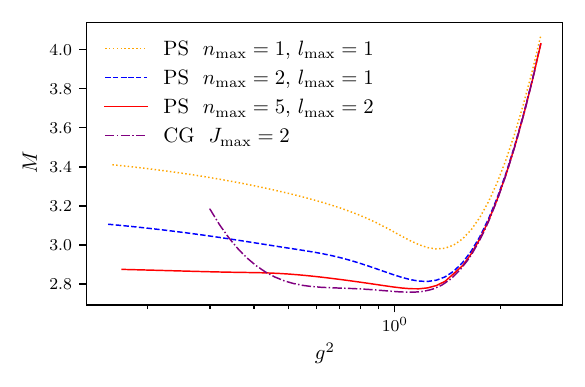}
    \caption{The mass gap $M$ for the system and operators plotted in
        \cref{fig:su2-gndState}.}
    \label{fig:su2-massGap}
\end{figure}

The picture is less clear, when studying the mass gap. This is shown in
\cref{fig:su2-massGap}. Again we do not have a GFMC result to compare to.
However, the rather large electric operators are expected to be reasonably
accurate up to the point where they diverge at around $g^2 < 0.6$. We can
clearly see that the plaquette state operators seem to converge towards the
electric results at stronger coupling. This leaves us hopeful, that this
approach will also deliver accurate results for SU(2) simulations at weak
couplings.
 
\section{Discussion and Conclusion}
\label{sec:conclusion}

The results presented in this work show that a functional basis constructed
from the eigenstates of the single-plaquette Hamiltonian provides an
efficient representation for Hamiltonian lattice gauge theories.
The advantage of this approach, named the plaquette-state basis, is that
it maintains a high accuracy at small values of the gauge coupling compared
to the other truncation schemes tested.   

For the Abelian U$(1)$ theory, the plaquette-state basis exhibits a consistent
and substantial reduction in error compared to both electric and magnetic bases
at a fixed operator dimension. The improvement is particularly pronounced in
the weak-coupling regime, where our results agree with GFMC
calculations over the entire coupling range.
Another advantage of the plaquette-state formulation is its versatility
in accessing a wider range of observables. While GFMC
provides highly accurate ground-state quantities, it is difficult to extract
excited-state properties such as the mass gap with satisfactory precision.
Moreover, the plaquette-state basis
shows no significant deterioration in accuracy up to the largest lattices
simulated. These features suggest that the basis captures the relevant low-energy
subspace of the Hamiltonian, offering a compact yet faithful
digitisation for both tensor-network and quantum-computing applications.

The plaquette state basis appears to retain efficiency in three
spatial dimensions, as our exploratory simulations in a $2\times 3^2$
volume indicate. Here we are currently working on a GPU implementation
which could enable simulations of larger three dimensional systems.

Preliminary investigations for the SU(2) case indicate that the same construction
extends to non-Abelian gauge theories. The plaquette-state basis reproduces
the available results for GFMC at strong and weak couplings keeping a higher versatility.
An open challenge lies in formulating a local dual Hamiltonian for non-Abelian gauge theories.
The lack of such a representation currently restricts simulations to relatively small lattices.
A local formulation would enable the application of tensor-network or quantum-computing
algorithms to larger systems while retaining the advantages of the plaquette-state basis.
Addressing this issue represents an important direction for future work.

It is worth mentioning that Green's function Monte Carlo
represents a valuable tool which we used to verify our
results for parameter values where no other method was
applicable. It remains to be seen whether GFMC can also be
applied for real time evolution or with a $\theta$-term or
baryon chemical potential.

The idea of constructing plaquette-based basis functions has recently attracted growing
attention. During this work we became aware of the independent development of a closely
related approach by P.~Fontana and collaborators, presented in 
Refs.~\cite{Miranda-Riaza:2025fus, Fontana:2024rux}.
Both projects appear to have originated from the insights of Bauer et al.~\cite{Bauer:2023jvw},
whose analysis provided many of the conceptual ingredients for the present method.
Here we presented complementary results and insights, in particular
more observables, three-dimensional simulations, and a comparison with other 
truncation schemes and GFMC.

The question of a suitable Hamiltonian for the non-Abelian case
remains. At the time of writing no suitable local formulation is known. This is
likely a necessity in order to extract non-trivial physical observables for larger scale
Hamiltonian simulations.

\begin{acknowledgments}
  This project was funded by the Deutsche Forschungsgemeinschaft (DFG,
  German Research Foundation) as a project in the CRC 1639 NuMeriQS --
  project no.\ 511713970. 
We thank ECT* for support at the Workshop \enquote{Hamiltonian
    Lattice Gauge Theories: Status, Novel Developments and
    Applications} during which this work has been developed.
This work is supported by the European
  Union’s Horizon Europe Framework Programme (HORIZON) under the ERA Chair
  scheme with grant agreement no. 101087126.
This work is supported with funds from the Ministry of Science,
  Research and Culture of the State of Brandenburg within the Centre
  for Quantum Technologies and Applications (CQTA).
  \flushright{\includegraphics[width=0.08\textwidth]{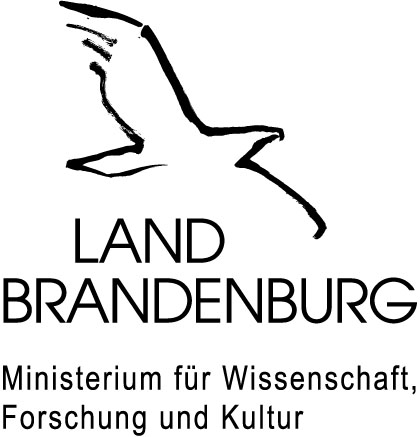}}
\end{acknowledgments}

\begin{appendix}
  \section{Green's Function Monte Carlo for Hamiltonian Lattice Gauge Theories}
\label{apdx:gfmc}

\begin{table*}
    \[
        \begin{array}{c|c|c|c|c|c}
            \text{Gauge Group}                                    &
            \tau_c                                                &
            E_L                                                   &
            F_{Q\,\mathbf{x},i}^c                                 &
            \sigma_{\chi}                                         &
            E_{\textrm{physical}}
            \\
            \hline
            \hline
            \text{U(1)}                                           &
            1                                                     &
            \displaystyle \begin{split} \left(2 \alpha -	\frac{2}{g^4}
                \right)
                & \sum_P \cos \varphi_P\\
                &- \frac{1}{4} \sum_{\mathbf{x},i}
                F_{Q\,\mathbf{x},i}^2 \end{split}
                                                                  &
            \displaystyle \alpha \sum_{\mathbf{x},i} \sum_{P \ni
                U_{\mathbf{x}, i} }
            s_P \left( U_{\mathbf{x}, i} \right) \sin \varphi_{P} &
            \displaystyle \sqrt{2\Delta t}                        &
            \displaystyle \frac{g^2}{2} \left(E_L + \frac{2}{g^2}
            N_{\textrm{plaquettes}} \right)
            \\
            \hline
            \text{SU(2)}                                          &
            \frac{1}{2} \sigma_c                                  &
            \displaystyle \begin{split} \left( \frac{3}{4} \alpha -
                \frac{2}{g^4}  \right) & \sum_P \mathrm{Tr}\left[ P \right] \\
                & - \frac{1}{2} \sum_{\mathbf{x},i,c}
                \left(F_{Q\,\mathbf{x},i}^c\right)^2 \end{split}
                                                                  &
\displaystyle \begin{split} \alpha	\sum_{\mathbf{x}, i, c} &
                \sum_{P \ni U_{\mathbf{x}, i} }
                \vec{x}_P (U_{\mathbf{x}, i} )
                ,  \\
                & \vec{x}_P (U_1)  = -\vec{x}_P (U_4) = \frac{1}{2} \mathrm{Tr}
                \left[\vec{\sigma} \, P \right] \\
                & \vec{x}_P (U_2) = \frac{1}{2} \mathcal{R}(U_1) \mathrm{Tr}
                \left[\vec{\sigma} \, P \right] \\
                & \vec{x}_P (U_3) = -\frac{1}{2} \mathcal{R}(U_4) \mathrm{Tr}
                \left[\vec{\sigma} \, P \right]
            \end{split}               &
            \sqrt{\Delta t}                                       &
            \displaystyle g^2 \left(E_L + \frac{4}{g^2} N_{\textrm{plaquettes}}
            \right)
            \\
        \end{array}
    \]
    \caption{Gauge group specific formulas for the implementation of a Green's
        Function Monte Carlo Solver. For U(1) the factor $s_P \left(
            U_{\mathbf{x}, i}
            \right)$ is equal to $+1$ when $P$ contains $U_{\mathbf{x}, i}$ and
        $-1$ when it contains
        $U_{\mathbf{x}, i}^\dagger$. In the SU(2) formulas we use the rotation
        matrices $\mathcal{R}^a_b (U) = 2 \,
            \mathrm{Tr} [U \tau^a U^\dagger \tau^b]$.
    }
    \label{tab:gfmcFormulas}
\end{table*}

To validate our simulations using digitised Hamiltonians, we compare to
Hamiltonian simulations using Green's Function Monte Carlo Methods. Here one
simulates the wave function $\psi(\mathbf{x})$ as an ensemble of random
walkers $\{\dots,\mathbf{x}_i, \dots\}$ in configuration space. The density of
walkers around a point $\mathbf{x}$ then is understood as the value of
$\psi(\mathbf{x})$. The ground state of a given system can then be found by
performing a probabilistic imaginary time evolution. These methods have a rich
history in nuclear physics \cite{PhysRev.128.1791,PhysRevC.36.2026}, but can
also be applied to lattice gauge theories.

Our implementation is derived from references
\cite{Chin1985,Hamer1996,Hamer2000}. In particular, we aim to calculate the
ground state energy and the ground state plaquette expectation value.
Ref.~\cite{Hamer2000} also suggests a method for estimating mass gaps. However,
we did not yet reach satisfactory precision in our results. In the
following we will give a quick overview of our implementation. For further
details we recommend the cited references.

For lattice gauge theories it is common to not actually simulate the wave
function itself, but the product of $\psi(\mathbf{x}) \phi_{\textrm{Trial}}
    (\mathbf{x})$, where $\phi_{\textrm{Trial}}$ is typically a variational
estimate of the ground state wave function. If chosen well, this can
significantly reduce the variance of measured observables.

As a trial wave function we use
\begin{equation}
    \phi_{\textrm{Trial}} = \frac{1}{N} \prod_{P}
    \mathrm{e}^{\alpha \mathrm{Tr} \left[ \mathrm{Re}(P) \right]}
\end{equation}
where $\alpha$ is chosen such that $\bra{\phi_{\textrm{Trial}}} \hat{H}
    \ket{\phi_{\textrm{Trial}}}$ is minimised.
The algorithm itself can then be summarised by the following pseudo-code:

\begin{algorithm}
    \caption{Green's Function Monte Carlo. Input parameters of the algorithm
        for are the coupling $g^2$, the step size $\Delta t$, the trial energy $E_T$,
        the number of sweeps $N_{\mathrm{sweeps}}$ and the number of propagation steps
        per branching step $N_{\mathrm{hits}}$. The gauge group specific formulas for $\tau_c, E_L, F_{Q\,
                    \mathbf{x},i}^c$ and $\sigma_\chi$ can be found in tab.~\ref{tab:gfmcFormulas}}
    \label{alg:gfmc}
    \begin{algorithmic}
        \Function{gfmc}{$g^2$, $\Delta t$, $E_T$,
            $N_{\textrm{sweeps}}$, $N_{\textrm{hits}}$} \\

        \State $\Psi \gets \left\{ C^1 = \{\dots, U^1_{\mathbf{x},i},\dots \},
            \dots,\right.$ \\
        $\left. \qquad \qquad \qquad \qquad C^{N_{\textrm{target}}} = \{\dots,
            U^{N_{\textrm{target}}}_{\mathbf{x},i},\dots \} \right\}$ \\
        \Comment{Random initial ensemble of gauge configurations} \\

        \State $\vec{w} \gets (1, \dots, 1)^T$ \\

        \For{$s = 1, \dots, N_{\textrm{sweeps}}$}

        \For{$C_j \in \Psi$} \Comment{Loop over
            configurations}
        \For{$U_{\mathbf{x},i} \in C$} \Comment{Loop over links} \\

        \For{$c=1,\dots,N_c$}
        \State $\chi^c \gets \textrm{random\_normal}(\mu=0,
            \sigma=\sigma_{\chi})$ \\
        \EndFor

        \State $W_{\textrm{rand}} \gets \exp \left( - \sum_c \chi^c  \tau_c
            \right)$ \\

        \Comment{Random component of the drift step} \\

        \State $W_{\textrm{force}} \gets \exp \left( - \Delta t \sum_c
            \,
            F_{Q\, \mathbf{x},i}^c (C_j) \tau_c \right)$ \\

        \Comment{Force component of the drift step} \\

        \State $U_{\mathbf{x},i} \gets W_{\textrm{rand}} \cdot
            W_{\textrm{force}} \cdot U_{\mathbf{x},i}$

        \Comment{Apply the drift Step}
        \EndFor

        \State $w_j \gets w_j \exp\left( \Delta t (E_L - E_T) \right)$ \\
        \Comment{Update the weights} \\
        \EndFor
        \\
        \If{$s \mod N_{\textrm{hits}} = 0$} \Comment{Branching Step}
        \\
        \State $\Psi \gets \Psi'$ s.t. $w_j$ copies of $C_j$ are in $\Psi'$ \\

        \State $\vec{w} \gets (1, \dots, 1)^T$ \Comment{Reset weights}
        \EndIf

        \EndFor

        \EndFunction
\end{algorithmic}
\end{algorithm}

After initialising the ensemble of gauge configurations $\Psi$ and their
respective weights $\vec{w}$, we enter the main loop of the algorithm.
First is the drift step. Here the probabilistic imaginary time evolution is
carried out.
For this, each link is rotated by the sum of a quantum force term denoted as
$F_Q^c$ originating from the use of the trial wave function and a random number
$\chi_c$ drawn from a normal distribution with variance
$\sigma_\chi^2$. Next, the weight of each walker is adjusted according to the
new energy of its configuration.

Lastly a branching step is performed every $N_{\textrm{hit}}$ steps. Here a new
ensemble $\Psi'$ is created from the old ensemble $\Psi$ by duplicating and
deleting the random walkers in $\Psi$ according to their weight. This ensures
that the ensemble does not contain too many walkers with vanishing weights and
thus implements a form of importance sampling. The fractional
part in a given weight is treated probabilistically. E.g. a walker with a
weight of $w_i = 2.1$ has a 10\% chance of appearing thrice in the new ensemble
and a 90\% chance of appearing twice.

The gauge group specific formulas for the energy $E_L$ and the quantum force
$F_Q^c$ are summarised in \cref{tab:gfmcFormulas}, together with the other
formulas needed in the implementation. As typically a renormalized Hamiltonian
$\hat{\widetilde{H}} \sim \hat{H} / g^2$ is simulated, the last column gives
conversion formulas to obtain the spectrum of $\hat{H}$ as given in
\cref{sec:theory} from the configuration energy $E_L$.

\subsection{Population Control}

\begin{figure*}
    \begin{subfigure}[t]{0.48\textwidth}
        \includegraphics[width=0.95\columnwidth]{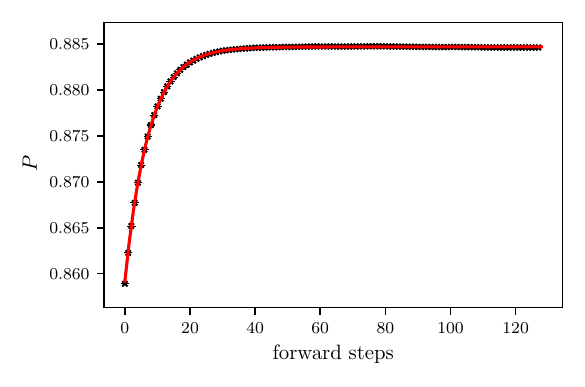}
        \caption{The forward walking plaquette estimator as a
            function of the forward steps at $\Delta t = 0.002 / \alpha$. Shown
            in red is the fit eq.~\eqref{eq:gfmcPlaqFit} used to estimate the
            plateau value.}
        \label{fig:gfmcPlaqFwd}
    \end{subfigure}
    \hfill
    \begin{subfigure}[t]{0.48\textwidth}
        \includegraphics[page=2,width=0.95\columnwidth]{plots/gfmcDtToZero}
        \caption{The ground state plaquette expectation value as a
            function of $\Delta t / \alpha$. In red, we show an extrapolation
            to $\Delta t \rightarrow 0$ which is used to estimate the physical
            value. }
        \label{fig:gfmcTtoZero}
    \end{subfigure}
    \caption{Representative Extrapolations for the extraction of observables
        from the GFMC data. This particular set of plots is obtained from a
        $3^3$
        lattice with open
        boundary conditions at $g^2=0.2645$ using U(1) as a gauge group.}
    \label{fig:gfmcExtrapolations}
\end{figure*}

The branching step deserves a little more attention, as it introduces a
problem. The trial energy $E_T$ will need continuous tuning, as otherwise the
population of random walkers will either grow or decay exponentially. Both
cases are obviously not acceptable for practical applications.

Thus, some form of population control is needed. As discussed in
\cite{PhysRevB.103.155135},
tuning $E_T$ dynamically introduces a small bias to the results. For most
method's this bias vanishes in the limit of infinitely large populations.
As the sensitivity to $E_T$ is quite high at weak couplings, a rather brute
force version of population control was used. At the beginning of each
branching step the weights are renormalized s.t.
\begin{equation}
    \tilde{w}_i = \frac{N_{\textrm{target}}}{\sum_j w_j} w_j \, .
\end{equation}
Then $N_{\textrm{target}}$ walkers were drawn from the old ones according to
the renormalized weights $\tilde{w}_i$.

In our runs we use $N_{\textrm{target}} = 500 \, 000$. Thus, we expect the bias
caused by population control to be quite small. As discussed earlier, good
matching with other simulation methods is achieved up to 5 significant digits.
However, it is likely that this bias needs to be investigated and addressed if
one wants to achieve higher precision.

\subsection{Observables}

The ground state energy can be measured in two different ways. First one can
take a weighted average of the configuration energies $E_L$ after every sweep.
This is the method we use will use in the following. Another option is to
estimate the trial energy $E_T$, that keeps the size of the ensemble constant.
Both should agree in the limit $\Delta t \rightarrow 0$.

To measure the plaquette expectation value we use the forward walking method as
proposed in \cite{Hamer2000}. For this one first measures the average plaquette
expectation value of each walker in a given ensemble. Then the ensemble is
propagated like discussed before. However, during the branching steps we keep
track of the ancestry of each walker. After a sufficient number of propagation
steps we estimate the plaquette as the average of the plaquette expectation
values of the initial ensemble, weighted by the weights of each configuration's
descendants:
\begin{equation}
    P = \frac{1}{\sum_i w^i_{\textrm{desc.}}} \sum_{i}	w^i_{\textrm{desc.}}
    P^i_{\textrm{anc.}}
\end{equation}
For a sufficient number of propagation steps this value eventually stabilises.
An example of this can be seen in \cref{fig:gfmcPlaqFwd}. To extract the
stabilised value a function of the form
\begin{equation}
    P(n) = P_\infty + b \, \mathrm{e}^{-c n}
    \label{eq:gfmcPlaqFit}
\end{equation}
is fitted to the data. $P_\infty $ is then taken as the estimator of the ground
state plaquette expectation value.

\subsection{Statistics}

As stated earlier the ensemble size was chosen to be  $500\, 000$. Branching
was performed every $N_{\textrm{hit}} = 5$ iterations. After a total $5 \times
    10\,000$ thermalisation iterations, initially at larger values of $\Delta
    t$, our measurement loop begins. It consists of $5 \times 64$ iterations to
estimate the energy, followed by the forward walking plaquette measurement
taken over $5 \times 128$ steps. $P(n)$ is measured after each branching step
in the forward propagation. These energy and plaquette measurements are then
repeated a few hundred times. In total measurements are thus taken over more
than $\mathcal{O}(5\times 20\,000)$ iterations.

For each cycle of the measurement loop the plaquette fit is carried out. The
energies are estimated by taking the mean of $E_L$ for the $5 \times 64$
initial sweeps. Errors are estimated by bootstrapping over all iterations of
the measurement loop.

\subsection{Extrapolating $\Delta t \rightarrow 0$}

Lastly, each observable must be extrapolated to the limit $\Delta t \rightarrow
    0$. As an example we show the plaquette expectation value as a function of
$\Delta t / \alpha$ in \cref{fig:gfmcTtoZero}. As can be seen, this is well
described by a linear fit. The intersection with $\Delta t = 0$ finally gives
us the values we used to compare to the digitised Hamiltonian results.

In our implementation we ran simulations at using four different time step
sizes $\Delta t$ given by
\begin{equation}
    \Delta t \in \{ 0.008 \, \alpha, \, 0.004 \, \alpha, \, 0.002 \, \alpha, \,
    0.001 \alpha \} \, .
\end{equation}
Expressing $\Delta t$ as a function of $\alpha$ proved useful, as it keeps the
overall change by each drift step independent of the coupling.

\subsection{Outlook}

In general, the GFMC results proved quite useful in testing and benchmarking
digitised Hamiltonian methods. We believe that it is worth reevaluating its
potential capabilities considering the renewed interest in Hamiltonian
lattice gauge theories as well as the advancements in computing power.
 \end{appendix}

\end{document}